\documentclass[12pt]{article}

\usepackage[letterpaper,top=2cm,bottom=2cm,left=2cm,right=2cm,marginparwidth=1.75cm]{geometry}
\usepackage{amsmath,amssymb}
\usepackage{graphicx}
\usepackage{subcaption}
\usepackage[colorlinks=true, allcolors=blue]{hyperref}
\usepackage{authblk}
\usepackage{xcolor}
\usepackage{tabularray}
\usepackage{bbm}
\usepackage{multirow}
\usepackage{cite}
\UseTblrLibrary{booktabs}
\usepackage[normalem]{ulem}
\usepackage{float}

\def\mathswitchr#1{\relax\ifmmode{\mathrm{#1}}\else$\mathrm{#1}$\fi}
\def\mathswitch#1{\relax\ifmmode#1\else$#1$\fi}
\def\mm{\mathswitch{\mathcal{M}}}
\def\GDF{\mathswitchr{GDF}}
\def\AIC{\mathswitchr{AIC}}
\newcommand{\brc}[1]{\left(#1\right)}

\newcommand{\bmu}{\mathswitch{\boldsymbol{\mu}}}
\newcommand{\btheta}{\mathswitch{\boldsymbol{\theta}}}

\newcommand{\bI}{\mathswitchr{\boldsymbol{I}}}

\newcommand{\bY}{\mathswitch{\boldsymbol{Y}}}

\newcommand{\bx}{\mathswitch{\boldsymbol{x}}}
\newcommand{\ceil}[1]{\left\lceil#1\right\rceil}

\newcommand{\declare}[2]{\vspace{2em}\noindent{\fontsize{14}{14}\selectfont\textbf{#1}}{%
\par\vspace{3pt}{\fontsize{12}{14}\selectfont #2}\par}}

\providecommand{\keywords}[1]
{
  \small
  \noindent
  \textbf{Keywords:} #1
}

\title{\textbf{\Large Measuring Neural Network Complexity via Effective Degrees of Freedom}}
\author{Jia Zhou \footnote{\href{mailto:jiazhou@buffalo.edu}{jiazhou@buffalo.edu}} }
\author{Douglas Landsittel \footnote{\href{mailto:dplansit@buffalo.edu}{dplansit@buffalo.edu}}}
\affil{Department of Biostatistics, University at Buffalo, Buffalo, NY 14214, USA}
\date{}

\begin{document}
\maketitle

\begin{abstract}
\fontsize{12pt}{14pt}\selectfont
Quantifying the complexity of feed-forward neural networks (FFNNs) remains challenging due to their nonlinear, hierarchical structure and numerous parameters. We apply generalized degrees of freedom (GDF) to measure model complexity in FFNNs with binary outcomes, adapting the algorithm for discrete responses. We compare GDF with both the effective number of parameters derived via log-likelihood cross-validation and the null degrees of freedom of Landsittel \textit{et al.}
Through simulation studies and a real data analysis, we demonstrate that GDF provides a robust assessment of model complexity for neural network models, as it depends only on the sensitivity of fitted values to perturbations in the observed responses rather than on assumptions about the likelihood. In contrast, cross-validation-based estimates of model complexity and the null degrees of freedom rely on the correctness of the assumed likelihood and may exhibit substantial variability. We find that GDF, cross-validation-based measures, and null degrees of freedom yield similar assessments of model complexity only when the fitted model adequately represents the data-generating mechanism. These findings highlight GDF as a stable and broadly applicable measure of model complexity for neural networks in statistical modeling.


\vspace{5mm}
\end{abstract}

\section{Introduction}
\label{sec:intro}
Feed-forward neural networks (FFNNs) constitute one of the most fundamental and widely applied architectures in statistical learning and machine intelligence. Owing to their universal approximation property~\cite{cybenko1989approximation,funahashi1989approximate,hornik1989multilayer}, FFNNs can approximate any continuous function to arbitrary accuracy given sufficient hidden units, allowing them to capture complex, nonlinear dependencies among covariates beyond the reach of traditional parametric regression models. Compared with more advanced neural network architectures, FFNNs remain attractive due to their simplicity, interpretability, and tractability in estimation~\cite{kuo2019interpretable,mcinerney2023feedforward,sun2025simple}. These features make FFNNs a natural and tractable starting point for investigating the relationship between model flexibility and complexity in neural networks.

Despite their remarkable flexibility, quantifying the model complexity of neural networks remains a challenging problem. In traditional regression models, such as linear or generalized linear models, model complexity is often tied directly to the number of estimated parameters, providing a clear and interpretable measure of the model's degrees of freedom. In contrast, the flexibility of neural networks arises from their hierarchical, nonlinear structure and the interplay between numerous parameters, activation functions, and regularization schemes. As a result, increasing model flexibility, for instance, by adding hidden layers or units, does not translate straightforwardly into a proportional increase in effective model complexity. This makes it difficult to assess how complex a fitted neural network truly is, in terms of the number of parameters effectively used to capture the underlying data structure.

Several approaches have been proposed to quantify model complexity through the concept of effective degrees of freedom (EDF), which measure the amount of information a model extracts from the data rather than simply counting the number of parameters.
Hastie and Tibshirani~\cite{hastie1990generalized} calculated the EDF for smoothing methods as the trace of the smoother (hat) matrix, extending the classical notion of degrees of freedom in linear regression to more flexible nonparametric models. 
Ye~\cite{ye1998measuring} introduced the notion of generalized degrees of freedom (GDF) as a unified framework for assessing model complexity across general modeling procedures. The GDF is defined as the sensitivity of the fitted values to perturbations in the observed responses, thus capturing the model effective adaptability to data. 
In the context of neural networks, Moody~\cite{moody1991effective} proposed the use of an effective number of parameters based on the curvature of the error surface, offering an analytic approximation to the network’s capacity. However, the practical computation of such quantities often requires the Hessian matrix of the loss function, which is computationally expensive and numerically unstable for large or highly nonlinear networks.
Related notions of the effective number of parameters have also been developed in connection with model selection and cross-validation frameworks.
Spiegelhalter \textit{et al.}~\cite{spiegelhalter2002bayesian} formulated a Bayesian measure of model complexity, defining the effective number of parameters as the difference between the posterior mean deviance and the deviance evaluated at the posterior means, which underlies the Deviance Information Criterion (DIC).
Stone~\cite{stone1977asymptotic} established the asymptotic equivalence between the Akaike Information Criterion (AIC) and leave-one-out cross-validation, bridging likelihood-based and prediction-based approaches to model assessment.
More recently, Hauenstein \textit{et al.}~\cite{hauenstein2018computing} explicitly connected the effective number of parameters in AIC to the expected change in log-likelihood cross-validation, providing a unified interpretation of these measures as estimators of model complexity and data adaptivity.
Together, these frameworks offer data-driven and theoretically grounded ways to quantify model complexity that are directly comparable across different model classes.
Moreover, Thorson~\cite{thorson2024measuring} applied the conditional Akaike Information Criterion (cAIC)~\cite{vaida2005conditional} to estimate the EDF, analogous to the GDF, for assessing model complexity in hierarchical models, and demonstrated its close correspondence with widely used Bayesian information criteria.

Early work on measuring the complexity of neural networks focused primarily on their likelihood-based behavior. Landsittel \textit{et al.}~\cite{landsittel2002null} and Landsittel~\cite{landsittel2009estimating} investigated the null distributions of likelihood ratio statistics for FFNNs and examined their approximation by chi-square distributions. The degrees of freedom were defined as the mean of the likelihood ratio statistics under the null, which we refer to as the \textit{null degrees of freedom}. This quantity served as an operational measure of model complexity, reflecting the effective dimensionality of the fitted network. Although informative, these approaches rely on assumptions about the asymptotic behavior of the likelihood ratio statistics and may be sensitive to deviations from the regularity conditions underlying maximum likelihood theory.

In this work, we propose an alternative approach for quantifying model complexity in FFNNs with binary outcomes by employing the GDF framework introduced by Ye~\cite{ye1998measuring}, where we modify the algorithm for estimating the GDF to make it suitable for classification outcomes. We first compare GDF with both the cross-validation-based effective number of parameters and the null degrees of freedom under intercept-only data-generating mechanisms. We then focus on the comparison between GDF and the cross-validation-based effective number of parameters under intercept-only and data-driven models that reflect the true data-generating mechanism.
In these settings, the two measures become noticeably closer under the true model than under the intercept-only model, indicating that agreement between GDF and likelihood-based measures improves as the fitted model better captures the underlying structure in the data.
The real data analysis further reveals that this behavior persists beyond controlled simulations. In practice, GDF and the effective number of parameters may diverge substantially, with the discrepancy depending strongly on model specification. Agreement between the two measures is observed primarily when the fitted model provides an adequate description of the data, whereas under model misspecification the cross-validation-based measure becomes less stable while GDF remains well defined. 
Therefore, we argue that the GDF provides a stable and robust measure of model complexity, whereas the effective number of parameters derived from AIC tends to depend more strongly on the accuracy of maximum likelihood estimation. Together, these analyses provide new insights into the complexity behavior of neural networks from a statistical perspective and highlight the potential of GDF-based methods for studying model flexibility in high-dimensional learning systems.

The rest of the paper is organized as follows. Section~\ref{sec:methods} introduces the methodology of the study, providing an overview of the feed-forward neural network model in Section~\ref{sec:methods:FFNNs}, and methods of quantifying effective degrees of freedom in Section~\ref{sec:methods:EDF}. Section~\ref{sec:results} presents numerical results, including simulation studies in Section~\ref{sec:results:simulation} and real data analysis in Section~\ref{sec:results:real}. Section~\ref{sec:conclusions} concludes the paper.

\section{Methodology}
\label{sec:methods}
\subsection{Feed-Forward Neural Network Model}
\label{sec:methods:FFNNs}
Feed-forward neural networks (FFNNs), also known as multilayer perceptrons, consist of an input layer, one or more hidden layers, and an output layer, where each unit in a given layer is typically connected to all units in the subsequent layer. This study focuses on FFNNs with one hidden layer, which has been demonstrated to possess universal approximation capability for measurable functions when the number of hidden units is sufficiently large~\cite{ripley1994neural}.

Let $y_k$ be the $k$-th observation of the response vector $\bY=\brc{y_1,\ldots,y_n}^T$ (or outputs), and $\bx_k=\brc{x_{1k},\ldots,x_{pk}}^T$ denote the corresponding $p$-dimensional vector of covariates (or inputs). Then the prediction of $y_k$ under the FFNN model is given by~\cite{ripley1994neural} 
\begin{equation}
  \hat{y}_k=\phi_o\left[v_0+\sum_{j=1}^Hv_j\phi_h\brc{w_{j0}+\sum_{i=1}^pw_{ji}x_{ik}}\right],
  \label{eq:ffnn}
\end{equation}
where $\phi_h\brc{\cdot}$ and $\phi_o\brc{\cdot}$ are the activation functions for the hidden and output layers, respectively. For the hidden layer, the activation function $\phi_h\brc{\cdot}$ is typically taken as the logistic function $\mathrm{expit}\brc{\cdot}$ such that 
$$
\mathrm{expit}\brc{z}=\frac{e^{z}}{1+e^{z}}. 
$$
For the output layer, the activation function $\phi_o\brc{\cdot}$ usually takes the identity function for continuous outcomes and the logistic function for binary outcomes, which is the case in our study.
The parameter $w_{ji}$ denotes the weight connecting the $i$-th covariate to the $j$-th hidden unit, and $w_{j0}$ the intercept term; $v_j$ is the weight connecting the $j$-th hidden unit to the output, and $v_0$ the intercept term associated with the output. Let $\boldsymbol{w}_i=\brc{w_{1i},\ldots,w_{Hi}}$ be the $H$-dimensional vector denoting all of the connection weights from input layer to the hidden layer for $i=0,1,\ldots,p$, and $\boldsymbol{v}=\brc{v_0,\ldots,v_H}$ the $\brc{H+1}$-dimensional vector denoting all of connection weights from the hidden layer to the output layer. Then,
\begin{equation}
  \btheta=\brc{\boldsymbol{v},\boldsymbol{w}_0,\ldots,\boldsymbol{w}_p}^T
  \label{eq:param}
\end{equation}
represents the vector of all parameters in the FFNN model with dimension $D=\brc{p+2}H+1$.

A penalized fit criterion is used in the model fitting via back-propagation algorithm~\cite{chauvin2013backpropagation}, such that the estimates of the weights $\hat{\btheta}$ are obtained by minimizing the following term, known as ridge-penalized objective 
\begin{equation}
  E+\lambda\lVert\btheta\rVert_2^2, 
\end{equation}
where $\lambda$ is the weight decay parameter, $\lVert\cdot\rVert_2^2$ denotes the squared $L_2$ norm, and $E$ is the fit criterion that can take either the sum of squares that is referred to as least-squares criterion
\begin{equation*}
  E=\lVert\boldsymbol{y}-\hat{\boldsymbol{y}}\rVert_2^2,
  \label{eq:E:ls}
\end{equation*}
or the deviance of a conditional likelihood that is referred to as entropy criterion
\begin{equation*}
  E=\sum_{k=1}^n\left[y_k\log\frac{y_k}{\hat{y}_k}+\brc{1-y_k}\log\frac{1-y_k}{1-\hat{y}_k}\right].
  \label{eq:E:entropy}
\end{equation*}
Ripley~\cite{ripley1994neural:ridge} recommended setting $\lambda\approx10^{-4} - 10^{-2}$ when using the least-squares criterion, and $\lambda\approx0.01 - 0.1$ when using the entropy criterion.
These recommendations were motivated by a Bayesian interpretation, under which the weights are assigned a Gaussian prior proportional to $\exp(-\lambda\lVert\boldsymbol{\theta}\rVert_2^2)$.

\subsection{Effective Degrees of Freedom}
\label{sec:methods:EDF}
We consider three measures of effective degrees of freedom: the generalized degrees of freedom (GDF), the effective number of parameters via log-likelihood cross-validation, and the null degrees of freedom. The first two are introduced in Sections~\ref{sec:methods:EDF:GDF} and~\ref{sec:methods:EDF:ENP}. The null degrees of freedom are defined in Section~\ref{sec:results:simulation:1} through the likelihood ratio statistic (Equation (\ref{eq:LRT})), and are approximated by its mean under the null model.

\subsubsection{Generalized Degrees of Freedom}
\label{sec:methods:EDF:GDF}
The generalized degrees of freedom (GDF) proposed by Ye~\cite{ye1998measuring} was originally applied to normally distributed outcomes. Specifically, given $\bY=\brc{y_1,\ldots,y_n}^T\sim N\brc{\bmu,\sigma^2\bI_n}$, the GDF for a modeling procedure $\mm:~\bY\to\hat{\bY}$ is given by
\begin{equation}
  \GDF=\sum_{i=1}^nh_i^{\mm}\brc{\bmu},
  \label{eq:gdf}
\end{equation}
where
$$
h_i^{\mm}\brc{\bmu}=\frac{\partial\mathbb{E}_{\bmu}\brc{\hat{y}_i}}{\partial\mu_i}
=\lim_{\delta\to0}\mathbb{E}_{\bmu}\brc{\frac{\hat{y}_i'-\hat{y}_i}{\delta}}, 
$$
and $\hat{y}_i$ and $\hat{y}_i'$ denote the $i$-th component of the predictions based on the original and perturbed observations ($y_i\to y_i+\delta$), respectively, for $i=1,\ldots,n$. 

For continuous outcomes, the definition of $h_i^{\mm}(\bmu)$ is naturally obtained through infinitesimal perturbations of each observation. In contrast, for binary outcomes perturbations are inherently discrete, since each component can only switch from $0$ to $1$ (or vice versa). To extend the notion of GDF in this setting, one may approximate
$$
h_i^{\mm}\brc{\bmu}\approx\mathbb{E}_{\bmu}\brc{\frac{\hat{y}_i'-\hat{y}_i}{\Delta_i}},
$$
where $\Delta_i=\brc{-1}^{y_i}$ indicates the direction of the perturbation for the $i$-th component, and thus approximate the GDF by 
\begin{equation}
\GDF\approx\sum_{i=1}^n\mathbb{E}_{\bmu}\brc{\frac{\hat{y}_i'-\hat{y}_i}{\Delta_i}}
=\mathbb{E}_{\bmu}\brc{\sum_{i=1}^n\frac{\hat{y}_i'-\hat{y}_i}{\Delta_i}}.
\label{eq:gdf:approx}
\end{equation}

From the right-hand side of (\ref{eq:gdf:approx}), since neural networks involve inherent stochasticity in training and prediction~\cite{scardapane2017randomness,zhuang2022randomness,raste2022quantifying}, the expectation can be approximated via Monte Carlo simulation. Specifically, in each replicate $j=1,\ldots,N$, let $\pi_j$ denote a random permutation of the index set $\{1,\ldots,n\}$, corresponding to the order in which components are perturbed. The empirical GDF estimator is then 
\begin{equation}
  \widehat{\GDF}=\frac{1}{N}\sum_{j=1}^N\sum_{i_j\in\pi_j}\frac{\hat{y}_{i_j}'-\hat{y}_{i_j}}{\Delta_{i_j}}.
\end{equation}
By treating the Monte Carlo replicates as approximately independent, the law of large numbers ensures that $\widehat{\GDF}$ provides a stable approximation to the expectation, with accuracy improving as $N$ increases.

On the other hand, the middle expression in (\ref{eq:gdf:approx}) is the basis for the algorithm that was originally proposed by Ye~\cite{ye1998measuring} for normally distributed outcomes and involves fitting a linear regression for repeated perturbations of each observation, where the perturbations are generated for all data points at once in each replicate. Here, for binary outcomes that switch between 0 and 1, it is not feasible to perform such ``\textit{global}'' perturbations, and meanwhile maintain the signal in the original data. Instead, in each replicate, only $k\ll n$ components are inverted (i.e., $0\leftrightarrow1$) to resemble a ``\textit{localized}'' perturbation to the original data. The algorithm for computing the GDF estimator is outlined as follows.
\begin{enumerate}
\item[a.] At step $t$, randomly select $k\brc{\ll n}$ components in $\bY$ and flip their values ($0\leftrightarrow1$) to obtain $\bY_*^{(t)}$. Then fit the neural network model such that $\bY_*^{(t)}\to\hat{\bY}_*^{(t)}$.
  
\item[b.] Repeat the above procedure until all components in $\bY$ are selected once, i.e., at each time, the randomly selected $k$ components are those not yet flipped in previous steps. Flip all the remaining components in the last step if the number of the remainders are less than $k$. 
  
\item[c.] Repeat $N$ times for \textbf{Steps a - b} (with a total number of perturbations $m=N\cdot\ceil{n/k}$). The perturbed observations and their predicted values are stored column-wise, i.e., $\bY_*=\left[\bY_*^{(1)},\bY_*^{(2)},\ldots,\bY_*^{(m)}\right]$ and $\hat{\bY}_*=\left[\hat{\bY}_*^{(1)},\hat{\bY}_*^{(2)},\ldots,,\hat{\bY}_*^{(m)}\right]$.

\item[d.] Let the $i$-th row of $\bY_*$ and $\hat{\bY}_*$ be $\bY_{*i}$ and $\hat{\bY}_{*i}$. Fit a linear regression to $\hat{\bY}_{*i}$ (responses) on $\bY_{*i}$ (predictors) to obtain $\hat{h}_i$, such that $\widehat{\mathrm{GDF}}=\sum_{i=1}^n\hat{h}_i$. 
\end{enumerate}

Note that in \textbf{Step d}, we regress $\hat{\bY}_{*i}$ on $\bY_{*i}$ directly, rather than using differences $\hat{\bY}_{*i}-\hat{\bY}_i$ versus $\bY_{*i}-\bY_i$. The baseline $\bY_i$ is constant, and although $\hat{\bY}_i$ varies due to neural network stochasticity, it can be stabilized by replacing it with the mean prediction $\overline{\hat{\bY}_{*i}}$, following the plug-in method suggested in~\cite{hauenstein2018computing}. This approach is often referred to as the ``\textit{horizontal}'' method and has been argued to be more robust than the previously described Monte Carlo-based approach that is commonly known as the ``\textit{vertical}'' method~\cite{elder2003generalization}.

\subsubsection{Effective Number of Parameters via Cross-validation}
\label{sec:methods:EDF:ENP}
Alternatively, the effective degrees of freedom can be estimated through the asymptotic equivalence between the AIC and log-likelihood cross-validation~\cite{stone1977asymptotic}. Following the notation in~\cite{hauenstein2018computing}, the $K$-fold log-likelihood cross-validation (CV) is defined as 
\begin{equation}
  \ell_{CV}=\sum_{i=1}^K\log f_{\hat{\btheta}^{[-i]}}\brc{\boldsymbol{y}_i},
  \label{eq:cv}
\end{equation}
where $\hat{\btheta}^{[-i]}$ is the parameter estimate obtained from the training data with the $i$-th fold $\boldsymbol{y}_i$ left out. It approximates the non-constant term in the Kullback-Leibler (KL) distance~\cite{kullback1951information} between the true density of the observations ($f_{\btheta}$) and the density of the observations with estimated parameters under the candidate model ($f_{\hat{\btheta}}$), which is given by 
\begin{equation}
  D_{KL}=\int f_{\btheta}\brc{\boldsymbol{y}}\log\frac{f_{\btheta}\brc{\boldsymbol{y}}}{f_{\hat{\btheta}}\brc{\boldsymbol{y}}}d\boldsymbol{y}
  \propto-\int f_{\btheta}\brc{\boldsymbol{y}}\log f_{\hat{\btheta}}\brc{\boldsymbol{y}}d\boldsymbol{y}. 
\end{equation}

The AIC as an asymptotically unbiased estimator of the expected KL distance~\cite{akaike1974new}, then is linked to the cross-validated deviance in such a way that
\begin{equation}
  \AIC=-2\ell_m+2p\stackrel{asym}{\approx}-2\ell_{CV},
  \label{eq:AIC:lcv}
\end{equation}
where $\ell_m$ is the logarithm of the maximum likelihood of the model with $\hat{\btheta}$, and $p$ is the effective number of parameters. Thus, the effective number of parameters, as a measure of model complexity, can be estimated by 
\begin{equation}
  \hat{p}_{CV}=\ell_m-\ell_{CV}.
  \label{eq:eff_num}
\end{equation}

  


It should be noted that when computing the $\ell_{CV}$ defined in (\ref{eq:cv}), the $K$ folds of the dataset are constructed so that the prevalence within each fold is approximately equal to that of the full dataset (i.e., stratified $K$-fold CV). The algorithm proceeds as follows.

\begin{enumerate}
\item[a.] Split the original dataset into two subsets: $S_0 = \{x : y_i = 0\}$ and $S_1 = \{x : y_i = 1\}$. Randomly assign fold indices ($1,\ldots,K$) to each element in $S_0$ and $S_1$. Elements with the same assigned index are grouped into the same fold, ensuring that the prevalence within each fold is approximately equal to the overall prevalence.

\item[b.] For each fold $i = 1,\ldots,K$, fit the neural network model on the training data $\boldsymbol{y}\setminus\boldsymbol{y}_i$ and obtain predictions for the $i$-th fold $\boldsymbol{y}_i$ under the trained model. Calculate the corresponding log-likelihood and combine results across all $K$ folds according to (\ref{eq:cv}). Then compute the estimated effective number of parameters $\hat{p}_{CV}$ using (\ref{eq:eff_num}).

\item[c.] Repeat \textbf{Steps a -- b} $N$ times to obtain the mean and standard deviation.
\end{enumerate}

Two common choices for the number of folds $K$ are often considered. Setting $K=n$ corresponds to the original leave-one-out cross-validation, which requires training the network $n$ times for a sample of size $n$ and is therefore computationally intensive. Another concern with $K=n$ is that the prevalence of $0/1$ outcomes may not be preserved within each fold. Alternatively, taking a smaller $K$, such as $5$ or $10$, is a more practical choice~\cite{ripley2007pattern}.

\section{Results}
\label{sec:results}
We use the \texttt{nnet} package in \texttt{R} to fit FFNN models in both the simulation studies and real data analysis. The entropy fit criterion is applied by setting \texttt{entropy = TRUE}. The weight decay parameter $\lambda$ is specified via the argument \texttt{decay}, and values of $0.01$, $0.05$, and $0.1$ are examined in the numerical studies. The number of hidden units $H$ is controlled through the \texttt{size} option and varied across $2$, $5$, and $10$ in different tests. The maximum number of iterations is set to \texttt{maxit = 1000} to ensure convergence.

\subsection{Simulation Studies}
\label{sec:results:simulation}
\subsubsection{Intercept-only Model Under Null Hypothesis}
\label{sec:results:simulation:1}
In this part of simulations, the response and covariates are generated independently to represent an intercept-only model under the null hypothesis, ensuring that the covariates exert no influence on the outcome. Table~\ref{tab:simu:scenarios} summarizes the simulation scenarios. 
\begin{table}[htbp]
  \centering
  \caption{Simulation scenarios under null hypothesis of intercept-only model.}
  \label{tab:simu:scenarios}
  \begin{tabular}{l*{2}{p{4cm}}p{6cm}}
    \toprule
    &Scenario &Observations &Inputs \\
    \midrule
    \multirow{2}{*}{1} &\multirow{2}{*}{Binary inputs} &$y_k\stackrel{iid}{\sim}\mathrm{Bern}\brc{0.3}$ &$x_{ik}\stackrel{iid}{\sim}\mathrm{Bern}\brc{0.5}$ \\
    & &{\scriptsize$k=1,\ldots,n$} &{\scriptsize$i=1,\ldots,p;~k=1,\ldots,n$} \\
    \midrule
    \multirow{2}{*}{2} &\multirow{2}{*}{Continuous inputs} &$y_k\stackrel{iid}{\sim}\mathrm{Bern}\brc{0.3}$ &$x_{ik}\stackrel{iid}{\sim}N\brc{0,1}$ \\
    & &{\scriptsize$k=1,\ldots,n$} &{\scriptsize$i=1,\ldots,p;~k=1,\ldots,n$} \\
    \midrule
    \multirow{2}{*}{3} &\multirow{2}{*}{Mixed inputs} &$y_k\stackrel{iid}{\sim}\mathrm{Bern}\brc{0.3}$ &$x_{ik}\stackrel{iid}{\sim}\mathrm{Bern}\brc{0.5},~x_{jk}\stackrel{iid}{\sim}N\brc{0,1}$ \\
    & &{\scriptsize$k=1,\ldots,n$} &{\scriptsize$i=1,\ldots,m;~j=m+1,\ldots,p;~k=1,\ldots,n$} \\
    \bottomrule
  \end{tabular}
\end{table}

We first examine the agreement between the two empirical estimates of the GDF obtained via the vertical and horizontal methods and the effective number of parameters ($\hat{p}_{CV}$) estimated through cross-validation. We set the sample size $n=200$ and the number of inputs $p=3$ for \textbf{Scenario 1} and \textbf{Scenario 2}, and $m=2,~p=3$ for \textbf{Scenario 3}. The number of hidden units is fixed at $H=5$, and the decay parameter is set as $\lambda=0.01$. The number of inversions $k$ in the horizontal method is varied from $1$ to $20$ (corresponding to 0.5\% - 10\% inversion rate), and the number of folds $K$ in cross-validation from $5$ to $20$. The empirical GDF ($\widehat{\mathrm{GDF}}$) and the effective number of parameters ($\hat{p}_{CV}$) are computed using the algorithms described in Sections~\ref{sec:methods:EDF:GDF} - \ref{sec:methods:EDF:ENP}. Each simulation is replicated $N=100$ times to obtain the means and standard errors~\footnote{There are two sources of variability: i) \textit{within-data} variability from internal iterative or bootstrap procedures; and ii) \textit{between-data} variability from external replications of the simulated datasets.}. It should be noted that each method for evaluating $\widehat{\mathrm{GDF}}$ or $\hat{p}_{CV}$ uses $100$ internal replications by default. Combined with the $N=100$ external replications, this results in $10^4\cdot n/k$ and $10^4\cdot K$ model fittings for the GDF and cross-validation approaches, respectively. In paricular, for the vertical method with $n=200$ and $k=1$, this corresponds to $2\times10^6$ model fittings.

\begin{figure}[h!]
  \centering
  \begin{subfigure}[t]{0.32\textwidth}
    \centering
    \caption{\textbf{Scenario 1}}
    \label{fig:cmp:s1}
    \includegraphics[scale=0.37]{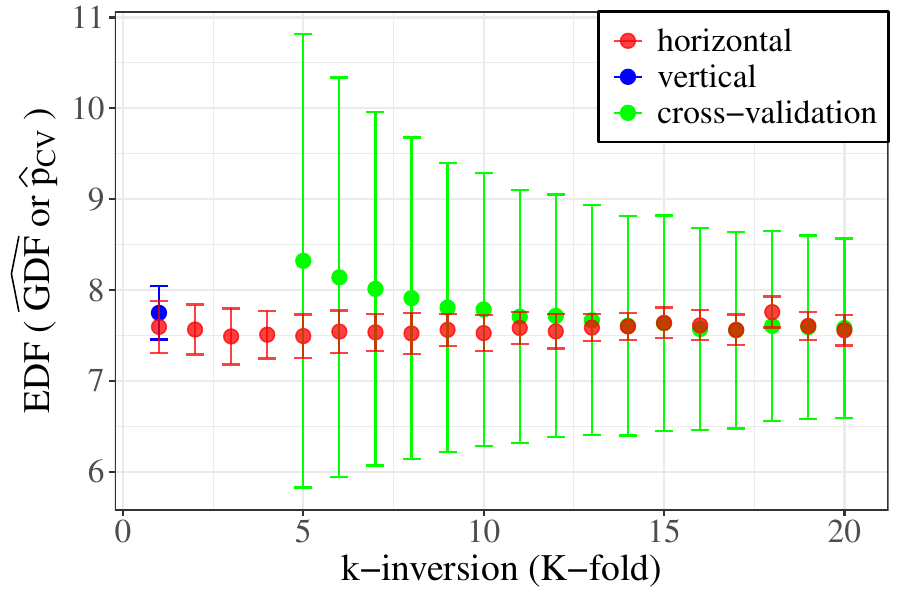}
  \end{subfigure}
  \begin{subfigure}[t]{0.32\textwidth}
    \centering
    \caption{\textbf{Scenario 2}}
    \label{fig:cmp:s2}
    \includegraphics[scale=0.37]{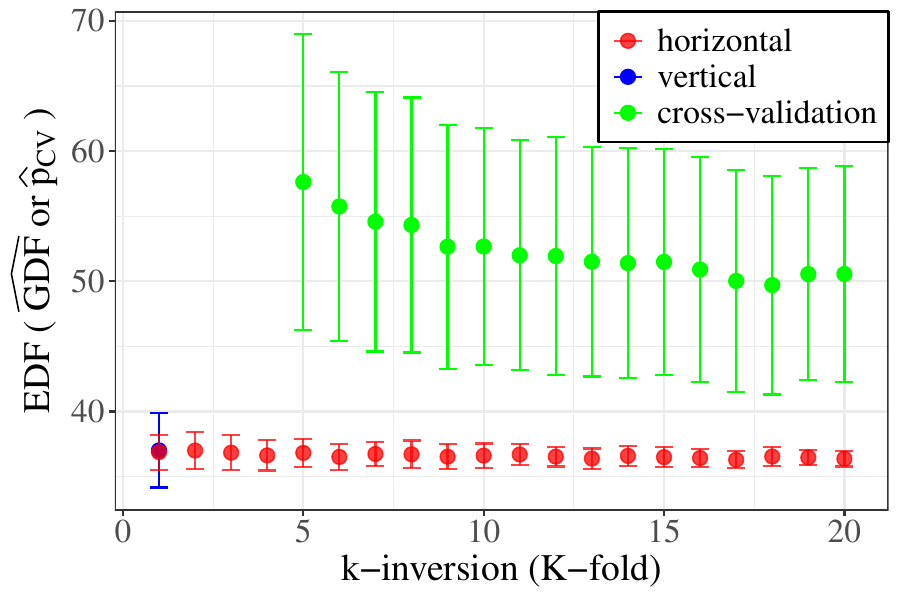}
  \end{subfigure}
  \begin{subfigure}[t]{0.32\textwidth}
    \centering
    \caption{\textbf{Scenario 3}}
    \label{fig:cmp:s3}
    \includegraphics[scale=0.37]{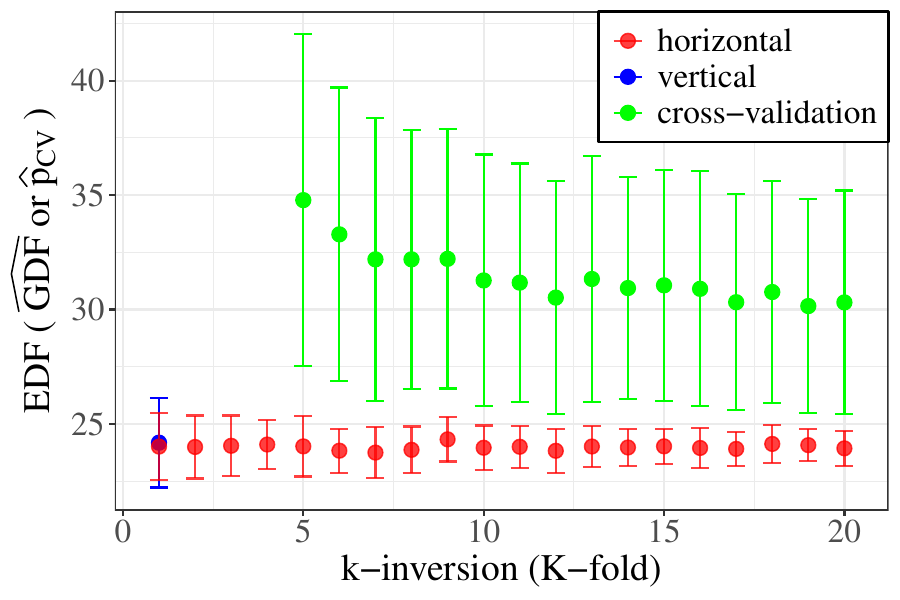}
  \end{subfigure}
  \caption{Comparison of vertical and horizontal GDF estimators ($\widehat{\mathrm{GDF}}$) and the cross-validation-based effective number of parameters ($\hat{p}_{CV}$) under three simulation scenarios (\textbf{Scenarios 1} - \textbf{3}). Horizontal $\widehat{\mathrm{GDF}}$ is evaluated for varying numbers of response inversions $k$, and $\hat{p}_{CV}$ is evaluated for varying numbers of folds $K$.}
  \label{fig:cmp}
\end{figure}

Figures \ref{fig:cmp:s1} - \ref{fig:cmp:s3} display the estimated effective degrees of freedom ($\widehat{\mathrm{GDF}}$ or $\hat{p}_{CV}$) by varying the number of inversions $k$ (for GDF) or the number of folds $K$ (for cross-validation) under \textbf{Scenarios 1} (three binary inputs), \textbf{Scenario 2} (three continuous inputs), and \textbf{Scenario 3} (a mix of two binary and one continuous input), respectively. Across all three scenarios, the two estimates of the GDF, which are obtained respectively using the vertical method with $k=1$ and the horizontal method for $k=1,\ldots,20$, show close agreement, indicating that both perturbation schemes provide consistent measures of model complexity. In \textbf{Scenario 1}, $\hat{p}_{CV}$ (computed for varying numbers of folds $K=5,\ldots,20$) aligns well with the GDF estimates. In contrast, in \textbf{Scenarios 2} and \textbf{3}, $\hat{p}_{CV}$ is substantially larger than the GDF. This inflation likely arises from the presence of continuous inputs, in such a way that, when continuous covariates are present, model fits can be more sensitive to training set variation across folds, leading to larger discrepancies between in-sample and cross-validated log-likelihoods and thus larger values of $\hat{p}_{CV}$. 

Regarding uncertainty, the error bars for the vertical GDF estimator are slightly wider than those of the horizontal estimator. This difference reflects the underlying algorithmic structures. The vertical method perturbs each response independently and refits the model a large number of times, inheriting substantial Monte Carlo variability. In contrast, the horizontal method fits a linear regression of predictions on perturbation magnitude to estimate the hat-matrix diagonals, which are then summed to obtain the GDF. This regression step stabilizes the estimate and yields lower variability even at $k=1$, with further reduction as $k$ increases. Finally, the error bars for $\hat{p}_{CV}$ are substantially larger than those of either GDF estimator, which is expected since $\hat{p}_{CV}$ aggregates variability from repeated data-splitting, model refitting within each fold, and stochastic weight initialization across multiple cross-validation runs. 

Next, we compare the empirical GDF, $\widehat{\mathrm{GDF}}$ (via horizontal method), and the effective number of parameters estimated by cross-validation, $\hat{p}_{CV}$, with the null degrees of freedom defined as the sample mean of the likelihood ratio statistic (LRT)~\cite{landsittel2002null,landsittel2009estimating}. Specifically, the likelihood ratio statistic is given by 
\begin{equation}
  \begin{aligned}
    \mathrm{LRT}&=2\brc{\ell_\text{fitted}-\ell_\text{null}} \\
    &=2\brc{\sum_{k=1}^n\brc{y_k\log\hat{y}_k+\brc{1-y_k}\log\brc{1-\hat{y}_k}}
    -n_1\log n_1-n_0\log n_0+n\log n}, 
  \end{aligned}
  \label{eq:LRT}
\end{equation}
where $y_k$ denotes the $k$-th observation, $\hat{y}_k$ the corresponding fitted value obtained from the FFNN model in Equation (\ref{eq:ffnn}), and $n_{1}$ ($n_0$) the count of ones (zeros) in observations such that 
$$
n_1=\sum_{k=1}^n y_k,
\quad
n_0=n-n_1. 
$$
The likelihood ratio statistic is postulated to follow an asymptotic chi-square distribution with degrees of freedom that reflect the effective model complexity. These degrees of freedom are therefore estimated by the sample mean of the statistic, denoted by $\overline{\mathrm{LRT}}$.

Table~\ref{tab:lrt:gdf:pcv} presents a detailed comparison among $\overline{\mathrm{LRT}}$, $\widehat{\mathrm{GDF}}$, and $\hat{p}_{CV}$ across a range of simulation settings for the three scenarios summarized in Table~\ref{tab:simu:scenarios}. Specifically, we vary the number of hidden units ($H=2,5,10$) and the decay parameter ($\lambda=0.01,0.05,0.1$) for models with sample size $n=200$ and $p=2,4,8$ inputs in \textbf{Scenarios 1} - \textbf{3}. In \textbf{Scenario 3}, half of the inputs are binary and half are continuous (i.e., $m=p/2$). For these comparisons, the number of inversions is fixed at $k=20$ ($10\%$ inversion rate) when computing $\widehat{\mathrm{GDF}}$ via horizontal method, and the number of folds is set to $K=10$ when estimating $\hat{p}_{CV}$ via cross-validation~\footnote{Unless otherwise specified, in all subsequent simulation studies and real data analyses the inversion rate for the horizontal GDF estimator is fixed at $10\%$ (i.e., $k=0.1n$, rounded to the nearest integer), and the number of folds for cross-validation is set to $K=10$.}.

{\scriptsize
\begin{longtblr}[
    caption={Comparison of null degrees of freedom estimates based on the likelihood ratio statistic ($\overline{\mathrm{LRT}}$), generalized degrees of freedom ($\widehat{\mathrm{GDF}}$, horizontal method), and effective number of parameters via cross-validation ($\hat{p}_{CV}$), where the numbers in parentheses are the standard errors.},
    label={tab:lrt:gdf:pcv}
  ]{
    colspec={*{2}{Q[c,4.5mm]}*{3}{Q[r]}Q[c]*{3}{Q[r]}Q[c]*{3}{Q[r]}}
  }
  \toprule
  \SetCell[r=2]{c}inputs &\SetCell[r=2]{c}$H$ &\SetCell[c=3]{c}$\lambda=0.01$ & & & &\SetCell[c=3]{c}$\lambda=0.05$ & & & &\SetCell[c=3]{c}$\lambda=0.1$ & & \\
  \cmidrule{3-5}\cmidrule{7-9}\cmidrule{11-13}
  & &$\overline{\mathrm{LRT}}$ &$\widehat{\mathrm{GDF}}$ &$\hat{p}_{CV}$ & &$\overline{\mathrm{LRT}}$ &$\widehat{\mathrm{GDF}}$ &$\hat{p}_{CV}$ & &$\overline{\mathrm{LRT}}$ &$\widehat{\mathrm{GDF}}$ &$\hat{p}_{CV}$ \\
  \midrule
  \SetCell[c=13]{c}\textbf{Scenario 1} & & & & & & & & & & & & \\
  \midrule
  2 &2 &3.14 &3.69(0.12) &3.18(0.88) & &2.71 &3.15(0.29) &2.57(0.97) & &2.25 &2.69(0.25) &1.95(0.76) \\
  &5 &3.17 &3.74(0.12) &3.22(0.87) & &2.76 &3.21(0.29) &2.56(0.95) & &2.33 &2.73(0.25) &1.99(0.73) \\
  &10 &3.07 &3.73(0.12) &3.22(0.88) & &2.63 &3.19(0.30) &2.51(0.92) & &2.17 &2.72(0.24) &1.94(0.73) \\
  4 &2 &12.53 &12.06(0.34) &13.80(2.79) & &9.92 &9.22(0.50) &9.14(2.05) & &6.85 &6.50(0.70) &5.89(1.98) \\
  &5 &16.00 &15.08(0.23) &18.52(2.77) & &12.97 &11.47(0.82) &11.98(2.80) & &7.32 &6.94(0.97) &6.44(2.28) \\
  &10 &16.14 &15.14(0.24) &18.79(2.84) & &12.96 &11.20(0.93) &11.62(2.83) & &7.32 &6.54(0.94) &6.06(2.22) \\
  8 &2 &37.26 &32.71(0.75) &54.88(11.68) & &28.87 &25.11(0.61) &30.69(5.35) & &22.85 &19.38(0.56) &20.85(3.49) \\
  &5 &101.08 &81.73(1.80) &181.11(24.80) & &81.48 &62.94(1.27) &88.68(11.58) & &61.45 &46.21(1.24) &53.39(6.54) \\
  &10 &157.24 &124.21(3.74) &272.36(39.81) & &125.45 &94.15(2.54) &134.39(18.70) & &79.93 &59.32(2.85) &67.30(10.09) \\
  \midrule
  \SetCell[c=13]{c}\textbf{Scenario 2} & & & & & & & & & & & & \\
  \midrule
  2 &2 &10.24 &10.16(0.39) &11.25(3.34) & &6.96 &6.91(0.47) &6.49(2.00) & &4.81 &5.02(0.58) &4.04(1.59) \\
  &5 &24.81 &22.44(0.51) &28.66(5.59) & &12.26 &11.32(0.91) &11.29(3.29) & &6.61 &6.38(0.86) &5.98(2.60) \\
  &10 &47.01 &39.06(1.03) &57.40(9.48) & &13.61 &12.12(1.31) &12.10(3.74) & &7.11 &6.62(1.10) &5.98(2.71) \\
  4 &2 &20.37 &19.31(0.49) &26.14(6.85) & &16.53 &15.13(0.45) &16.69(3.79) & &13.36 &12.13(0.45) &12.21(2.75) \\
  &5 &56.69 &49.74(0.89) &79.58(13.54) & &45.16 &37.45(0.72) &47.79(7.75) & &35.61 &28.18(0.95) &32.07(5.57) \\
  &10 &123.13 &100.03(2.09) &227.50(28.88) & &86.55 &66.12(1.21) &92.56(12.51) & &50.79 &39.08(2.42) &42.70(7.82) \\
  8 &2 &39.20 &36.08(0.67) &56.00(12.97) & &35.58 &31.40(0.69) &40.82(7.48) & &31.03 &26.84(0.62) &31.50(5.37) \\
  &5 &117.66 &99.22(1.53) &228.45(31.26) & &106.32 &84.00(0.99) &133.08(16.73) & &93.01 &70.72(0.80) &92.38(12.06) \\
  &10 &219.16 &178.43(0.95) &403.63(57.12) & &188.47 &143.29(1.07) &223.89(29.70) & &156.81 &116.42(1.18) &149.05(18.40) \\
  \midrule
  \SetCell[c=13]{c}\textbf{Scenario 3} & & & & & & & & & & & & \\
  \midrule
  2 &2 &8.37 &8.39(0.41) &9.24(2.59) & &5.16 &5.27(0.49) &4.81(1.76) & &3.48 &3.82(0.50) &3.17(1.43) \\
  &5 &15.16 &13.84(0.80) &15.92(3.62) & &6.03 &6.19(0.70) &5.63(2.23) & &3.84 &4.10(0.62) &3.41(1.64) \\
  &10 &17.19 &15.74(1.30) &18.60(4.83) & &6.30 &6.16(0.69) &6.06(2.38) & &3.89 &4.01(0.62) &3.34(1.58) \\
  4 &2 &18.98 &17.70(0.47) &23.67(6.11) & &14.13 &12.97(0.47) &14.26(3.32) & &10.77 &9.79(0.59) &9.74(2.35) \\
  &5 &51.15 &43.39(0.70) &68.92(11.46) & &35.80 &28.95(0.93) &34.43(5.36) & &22.50 &17.89(1.64) &18.23(4.91) \\
  &10 &102.60 &81.82(1.57) &170.83(22.87) & &52.65 &41.79(2.41) &51.71(8.84) & &22.89 &18.69(2.09) &19.37(5.53) \\
  8 &2 &39.37 &34.98(0.75) &58.13(13.45) & &33.05 &28.65(0.59) &36.78(6.68) & &27.46 &23.28(0.49) &26.50(4.76) \\
  &5 &113.24 &94.74(1.53) &216.02(30.63) & &96.93 &75.76(0.87) &114.49(15.26) & &79.79 &60.16(0.76) &74.47(9.73) \\
  &10 &214.22 &171.01(1.44) &381.04(52.10) & &171.34 &129.23(1.28) &194.11(24.80) & &133.35 &97.89(1.23) &119.66(14.24) \\
  \bottomrule
\end{longtblr}
}

As can be seen, $\overline{\mathrm{LRT}}$ tends to agree more closely with $\hat{p}_{CV}$ than with $\widehat{\mathrm{GDF}}$, in the sense that $\overline{\mathrm{LRT}}$ more frequently falls within the variability range of $\hat{p}_{CV}$ than that of $\widehat{\mathrm{GDF}}$. Nevertheless, $\widehat{\mathrm{GDF}}$ is often of comparable magnitude to $\overline{\mathrm{LRT}}$, indicating a reasonable level of agreement between these two measures. Noticeable discrepancies mainly arise in more complex network settings, characterized by larger numbers of inputs and hidden units and weaker regularization (small $\lambda$), where $\overline{\mathrm{LRT}}$ may lie outside the variability range of both $\widehat{\mathrm{GDF}}$ and $\hat{p}_{CV}$. 


A plausible explanation for this pattern is that both $\overline{\mathrm{LRT}}$ and $\hat{p}_{CV}$ are more directly tied to likelihood-based model fitting than $\widehat{\mathrm{GDF}}$. The null degrees of freedom based on the likelihood ratio statistic are defined through difference in log-likelihood between the fitted and null models, while $\hat{p}_{CV}$ is constructed from the discrepancy between in-sample and cross-validated log-likelihood. In contrast, the GDF is estimated through perturbations of the observed responses and reflects the sensitivity of predictions to such perturbations, capturing a broader notion of data adaptivity rather than likelihood-based complexity. As a result, $\overline{\mathrm{LRT}}$ and $\hat{p}_{CV}$ tend to show closer agreement, particularly when the network is relatively simple or strongly regularized, settings in which likelihood-based approximations are more stable. 

It is important to note that a key distinction between GDF and likelihood-based measures of model complexity lies in their domain of validity. The generalized degrees of freedom can always be computed for a fitted procedure, as it is defined through the sensitivity of fitted values to perturbations in the observed responses and does not rely on likelihood correctness or distributional assumptions. In contrast, likelihood-based quantities such as the likelihood ratio statistic, AIC-derived effective number of parameters, and cross-validation-based penalties are meaningful only when the fitted model provides an adequate description of the data-generating mechanism. Under model misspecification, these likelihood-based measures may become unreliable or lose their theoretical justification. This robustness to model misspecification constitutes a key advantage of GDF as a measure of model complexity for flexible models such as neural networks. 


\subsubsection{True Model versus Intercept-only Model}
\label{sec:results:simulation:2}
As been seen, under the null hypothesis, the resulting estimated GDF and $\hat{p}_{CV}$ have notable discrepancy (i.e., $\widehat{\mathrm{GDF}}<\hat{p}_{CV}$) under \textbf{Scenario 2} and \textbf{3}, which is largely driven by the continuous inputs in model fit. To further investigate these behaviors, we conduct simulations under two data-generating mechanisms:
\begin{itemize}
\item \textbf{True model generation}: the outcomes are generated as 
  $$
  y_k\sim\mathrm{Bern}\brc{\hat{y}_k}, 
  $$
  where $\hat{y}_k$ is the FFNN prediction in~\eqref{eq:ffnn}. The weights are initialized/generated as
  $$
  v_0=1,~\brc{v_1,\ldots,v_H}^T\sim N\brc{0,s\bI_H},~\boldsymbol{w}_i\sim N\brc{0,s\bI_H}, 
  $$
  with $s=1$, and the inputs are sampled as $x_{ik}\stackrel{iid}{\sim}N\brc{0,1}$, for $i=1,\ldots,p$. 

\item \textbf{Intercept-only model generation}: the outcomes follow
  $$
  y_k\stackrel{iid}{\sim}\mathrm{Bern}\brc{0.5}, 
  $$
  and the inputs are independently generated as $x_{ik}\stackrel{iid}{\sim}N\brc{0,1}$, analogous to \textbf{Scenario 2} but with a different Bernoulli success probability. 
\end{itemize}

We denote the GDF and $\hat{p}_{CV}$ estimated under the first mechanism by $\widehat{\mathrm{GDF}}^{true}$ and $\hat{p}_{CV}^{true}$, and those estimated under the second mechanism by $\widehat{\mathrm{GDF}}^{int}$ and $\hat{p}_{CV}^{int}$.

We expect the discrepancy between GDF and $\hat{p}_{CV}$ to diminish under the true model compared with the intercept-only model. The rationale is twofold. First, the estimated GDF tends to be larger under the true model, i.e., $\widehat{\mathrm{GDF}}^{true}>\widehat{\mathrm{GDF}}^{int}$, because the fitted values are more responsive to perturbations in the observed responses when the data are generated from the model structure being fitted. Second, the cross-validated log-likelihood is expected to improve under the true model due to better generalization, while the in-sample log-likelihood changes little between the two settings. This occurs because the network is always trained to maximize the likelihood on the observed data. Thus, even when the model is misspecified (as in the intercept-only case), the FFNN still adjusts its weights to fit the training outcomes closely, yielding similar in-sample likelihoods. In contrast, the cross-validated likelihood is more sensitive to misspecification, such that under the intercept-only setting, the network overfits noise, leading to poorer out-of-sample performance. Since $\hat{p}_{CV}$ is defined as the difference between the in-sample and cross-validated log-likelihood, this yields $\hat{p}_{CV}^{true}<\hat{p}_{CV}^{int}$. Overall, these opposing movements, i.e., larger GDF and smaller $\hat{p}_{CV}$ under the true model, narrow the discrepancy between the two measures. This gap is further reduced when the model structure is simpler (e.g., fewer inputs or smaller hidden layers) or when stronger regularization (larger decay parameter) is applied, both of which mitigate overfitting and stabilize the relationship between GDF and $\hat{p}_{CV}$.

\begin{figure}[htpb]
  \centering
  \includegraphics[scale=0.75]{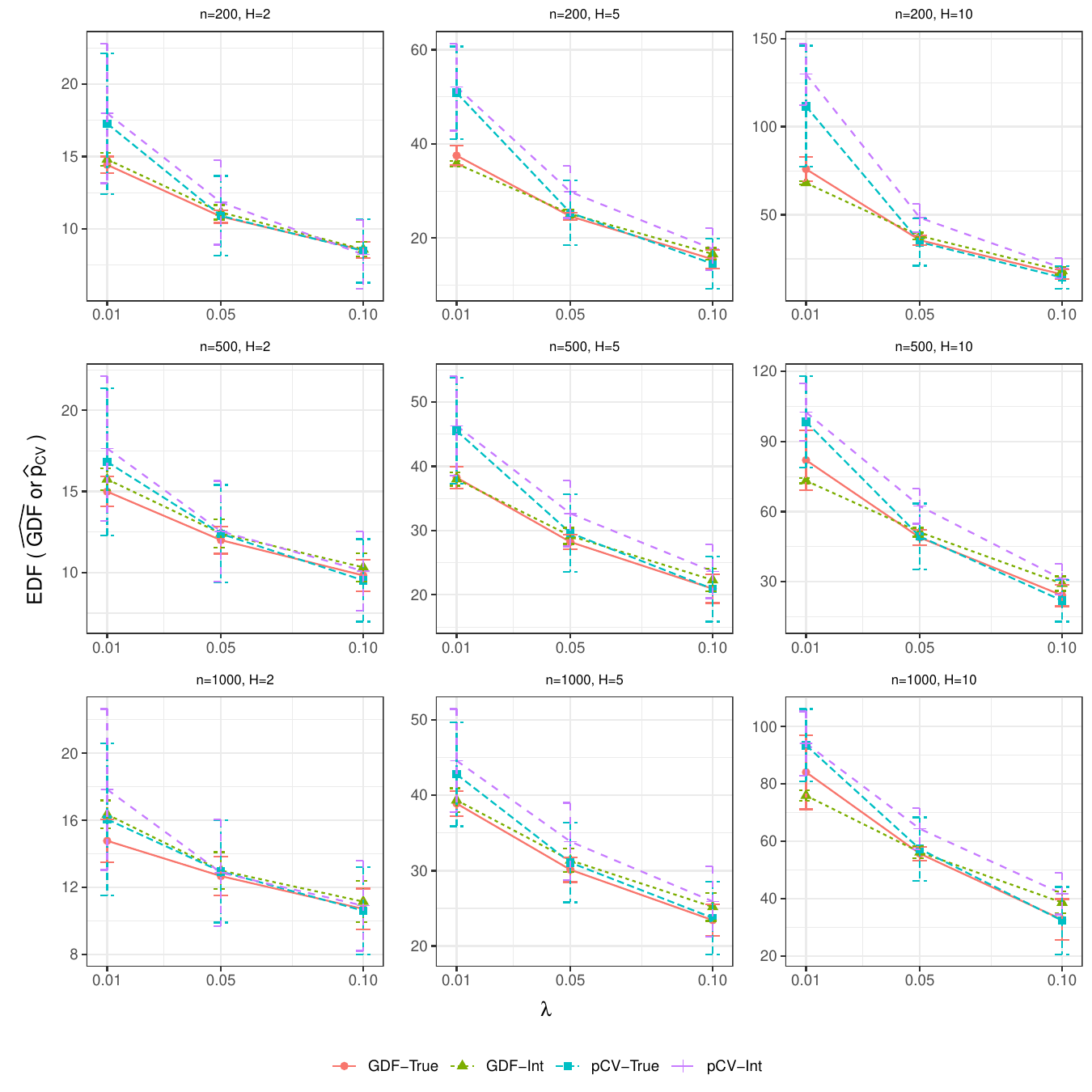}
  \caption{Generalized degrees of freedom (GDF) and effective number of parameters ($\hat{p}_{CV}$) under the true and intercept-only models for a neural network with three continuous inputs, where the red dotted curve the estimated GDF under the true model ($\widehat{\mathrm{GDF}}^{true}$), the green triangular curve the estimated GDF under the intercept-only model ($\widehat{\mathrm{GDF}}^{int}$), the cyan square curve the estimated $\hat{p}_{CV}$ under the true model ($\hat{p}_{CV}^{true}$), and the purple cross-marked curve the estimated $\hat{p}_{CV}$ under the intercept-only model ($\hat{p}_{CV}^{int}$). Panels are arranged by sample size ($n=200,~500,~1000$ from top to bottom) and number of hidden units ($H=2,~5,~10$ from left to right). Within each panel, results are shown for decay parameters $\lambda=0.01,~0.05,~0.1$.}
  \label{fig:true_vs_null:x3}
\end{figure}

\begin{figure}[htpb]
  \centering
  \includegraphics[scale=0.75]{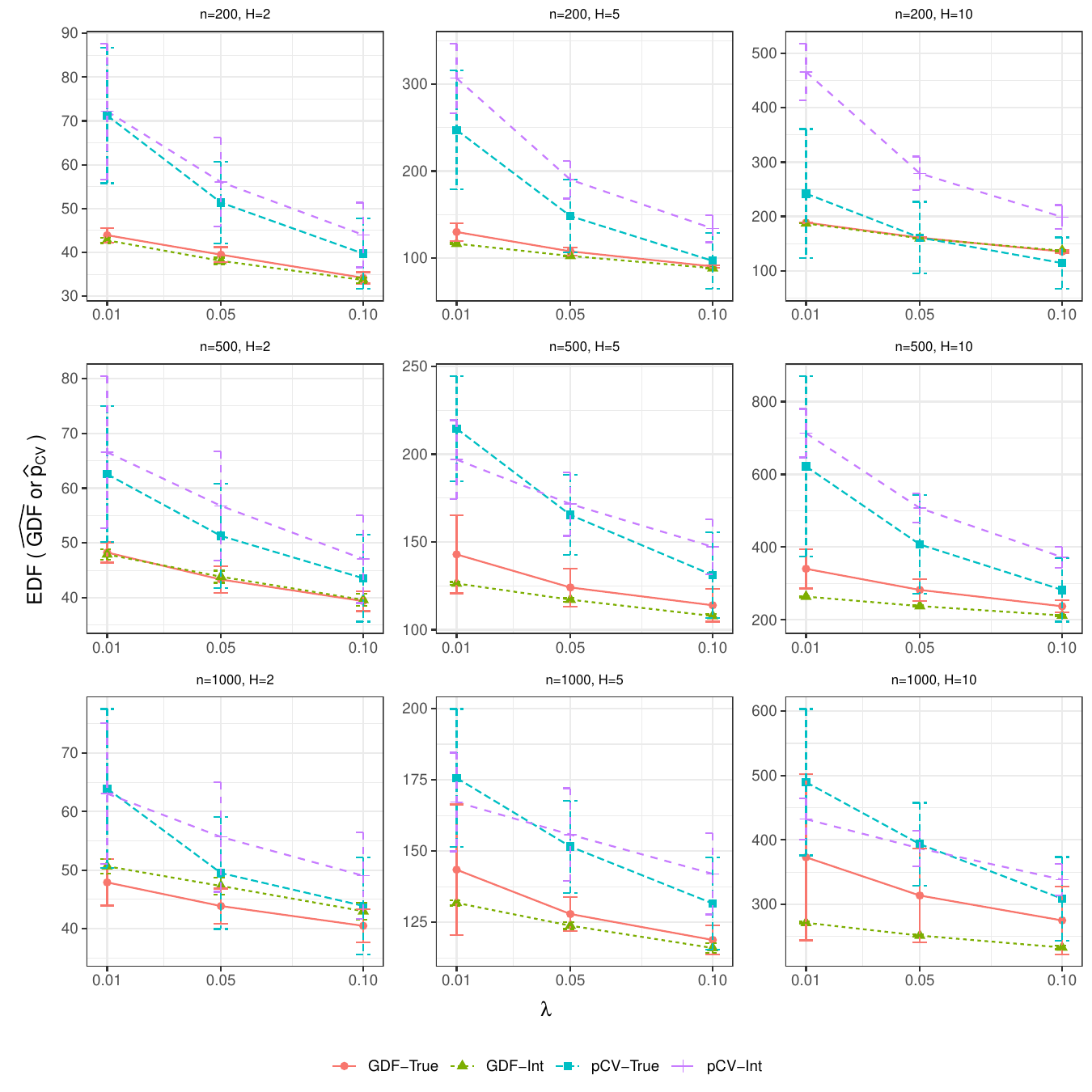}
  \caption{Same as Figure~\ref{fig:true_vs_null:x3}, but for a neural network with ten continuous inputs.}
  \label{fig:true_vs_null:x10}
\end{figure}
To illustrate these behaviors, Figures~\ref{fig:true_vs_null:x3} and~\ref{fig:true_vs_null:x10} present two sets of $3\times3$ panels comparing the estimated GDF and $\hat{p}_{CV}$ under the true and intercept-only models. Figure~\ref{fig:true_vs_null:x3} corresponds to the neural network with three continuous inputs, while Figure~\ref{fig:true_vs_null:x10} shows results for the network with ten continuous inputs. In both figures, panels are organized by sample size ($n=200,~500,~1000$ from top to bottom) and number of hidden units ($H=2,~5,~10$ from left to right). Within each panel, the decay parameter varies over $\lambda=0.01,~0.05,~0.1$. For each configuration, the curves display $\widehat{\mathrm{GDF}}^{true}$ (red dotted), $\widehat{\mathrm{GDF}}^{int}$ (green triangular), $\hat{p}_{CV}^{true}$ (cyan square), and $\hat{p}_{CV}^{int}$ (purple cross-marked), allowing direct visual comparison of the two complexity measures across different sample sizes, network architectures, and regularization strengths.

As can be seen, the expected ordering $\widehat{\mathrm{GDF}}^{true}>\widehat{\mathrm{GDF}}^{int}$ and $\hat{p}_{CV}^{true}<\hat{p}_{CV}^{int}$ becomes increasingly pronounced when the model is more flexible, namely, with smaller decay, larger numbers of hidden units, and more input variables. In contrast, when the model structure is relatively simple (e.g., $H=2$), these patterns are less distinct and may even occasionally reverse, though the ranges of the estimates for the true and intercept-only models largely overlap in these cases. 
A clear pattern also appears with respect to sample size. As $n$ increases, the estimated GDF generally increases slightly for each fixed configuration of $H$ and $\lambda$, since larger samples allow the fitted model to react more stably and detectably to perturbation in the responses. In contrast, $\hat{p}_{CV}$ generally decreases with larger $n$ because cross-validated log-likelihood improves more substantially than the in-sample log-likelihood. With GDF drifting upward and $\hat{p}_{CV}$ drifting downward, the two quanties usually move closer together, and thus the discrepancy between the estimated GDF and $\hat{p}_{CV}$ tends to diminish as the sample size $n$ grows, in spite of a few exceptions. Nevertheless, this sample size effect remains secondary compared to the influence of model complexity (inputs, hidden units) and the decay parameter.

To quantitatively assess the visual patterns observed in Figures~\ref{fig:true_vs_null:x3} and~\ref{fig:true_vs_null:x10}, we conduct paired-$t$ tests comparing the estimates obtained under the true and intercept-only models. The differences
$$
\Delta\widehat{\mathrm{GDF}}=\widehat{\mathrm{GDF}}^{true}-\widehat{\mathrm{GDF}}^{int}
\quad
\text{ and }
\quad
\Delta\hat{p}_{CV}=\hat{p}_{CV}^{true}-\hat{p}_{CV}^{int}
$$
are constructed within each simulation iteration in such a way that, for a given iteration, the GDF and $\hat{p}_{CV}$ under the true and intercept-only models are computed consecutively, and the random seed is reset before each model fit, ensuring that the initialization of network weights is identical across the two conditions. This design places the two estimates on the same footing and substantially reduces variability due to random initialization.

The resulting 95\% confidence intervals from the paired-$t$ tests are reported in Table~\ref{tab:diff:paired-t}. Overall, the statistical results align closely with the qualitative trends in the figures, i.e., $\Delta\widehat{\mathrm{GDF}}$ is generally positive and increases with model complexity or weaker regularization, whereas $\Delta\hat{p}_{CV}$ is generally negative and becomes more pronounced under the same conditions, with only a few minor exceptions. 
{\scriptsize
\begin{longtblr}[
    caption={95\% confidence intervals for the difference in GDF ($\Delta\widehat{\mathrm{GDF}}=\widehat{\mathrm{GDF}}^{true}-\widehat{\mathrm{GDF}}^{int}$) and $\hat{p}_{CV}$ ($\Delta\hat{p}_{CV}=\hat{p}_{CV}^{true}-\hat{p}_{CV}^{int}$) between true and intercept-only models, based on paired-$t$ test.},
    label={tab:diff:paired-t}
  ]{
    cell{1}{1}={r=2}{valign=m},
    cell{1}{2}={r=2}{valign=m},
    cell{1}{3}={r=2}{valign=m},
    cell{1}{4}={c=2}{halign=c},
    cell{1}{7}={c=2}{halign=c},
    cell{1}{10}={c=2}{halign=c},
    cell{3}{1}={r=6}{valign=h},
    cell{9}{1}={r=6}{valign=h},
    cell{15}{1}={r=6}{valign=h},
    cell{3}{2}={r=3}{valign=h},
    cell{6}{2}={r=3}{valign=h},
    cell{9}{2}={r=3}{valign=h},
    cell{12}{2}={r=3}{valign=h},
    cell{15}{2}={r=3}{valign=h},
    cell{18}{2}={r=3}{valign=h},
    colspec={Q[c]Q[1cm,c]Q[c]*{2}{Q[2cm,c]}Q[c]*{2}{Q[2cm,c]}Q[c]*{2}{Q[2cm,c]}}
  }
  \toprule
  $n$ &\# inputs &$H$ &$\lambda=0.01$ & & &$\lambda=0.05$ & & &$\lambda=0.10$ \\
  \cline{4-5}\cline{7-8}\cline{10-11}
   & & &$\Delta\widehat{\mathrm{GDF}}$ &$\Delta\hat{p}_{CV}$ & &$\Delta\widehat{\mathrm{GDF}}$ &$\Delta\hat{p}_{CV}$ & &$\Delta\widehat{\mathrm{GDF}}$ &$\Delta\hat{p}_{CV}$ \\
  \midrule
  200 &3 &2 &$-0.37\pm0.14$ &$-0.71\pm0.65$ & &$-0.29\pm0.14$ &$-0.93\pm0.50$ & &$-0.06\pm0.14$ &$0.26\pm0.41$ \\
  & &5 &$1.74\pm0.41$ &$-1.22\pm1.41$ & &$-0.53\pm0.21$ &$-4.44\pm1.46$ & &$-1.13\pm0.51$ &$-3.14\pm1.33$ \\
  & &10 &$7.86\pm1.40$ &$-18.23\pm6.11$ & &$-2.42\pm0.70$ &$-13.83\pm2.90$ & &$-2.05\pm0.65$ &$-5.38\pm1.48$ \\
  \midrule{2-11}
  &10 &2 &$1.17\pm0.35$ &$-0.95\pm1.70$ & &$1.42\pm0.39$ &$-4.69\pm1.52$ & &$0.51\pm0.30$ &$-4.26\pm1.33$ \\
  & &5 &$13.52\pm2.02$ &$-59.64\pm12.17$ & &$5.31\pm0.96$ &$-41.93\pm7.74$ & &$1.97\pm0.31$ &$-37.19\pm6.14$ \\
  & &10 &$1.60\pm0.14$ &$-223.99\pm23.44$ & &$1.06\pm0.14$ &$-118.13\pm13.76$ & &$-1.63\pm0.45$ &$-84.79\pm9.84$ \\
  \midrule
  500 &3 &2 &$-0.76\pm0.22$ &$-0.83\pm0.62$ & &$-0.41\pm0.23$ &$-0.15\pm0.50$ & &$-0.50\pm0.26$ &$-0.58\pm0.49$ \\
  & &5 &$0.25\pm0.40$ &$-0.78\pm0.98$ & &$-0.95\pm0.32$ &$-3.04\pm1.03$ & &$-1.35\pm0.52$ &$-2.73\pm0.96$ \\
  & &10 &$8.81\pm2.61$ &$-4.13\pm3.25$ & &$-2.27\pm0.69$ &$-12.99\pm2.76$ & &$-5.02\pm1.13$ &$-9.19\pm2.12$ \\
  \midrule{2-11}
  &10 &2 &$0.36\pm0.41$ &$-4.01\pm1.74$ & &$-0.51\pm0.53$ &$-5.48\pm1.69$ & &$-0.21\pm0.42$ &$-3.5\pm1.35$ \\
  & &5 &$16.58\pm4.39$ &$17.66\pm3.61$ & &$6.96\pm2.16$ &$-6.23\pm3.51$ & &$6.12\pm1.87$ &$-16.16\pm4.37$ \\
  & &10 &$76.51\pm10.77$ &$-91.36\pm47.88$ & &$44.45\pm5.89$ &$-100.06\pm26.2$ & &$25.32\pm3.35$ &$-90.07\pm16.8$ \\
  \midrule
  1000 &3 &2 &$-1.58\pm0.30$ &$-1.78\pm0.79$ & &$-0.33\pm0.31$ &$0.07\pm0.46$ & &$-0.43\pm0.35$ &$-0.30\pm0.43$ \\
  & &5 &$-0.44\pm0.43$ &$-1.81\pm0.91$ & &$-1.25\pm0.42$ &$-2.79\pm0.92$ & &$-1.76\pm0.56$ &$-2.19\pm0.96$ \\
  & &10 &$8.17\pm2.64$ &$-0.65\pm1.77$ & &$-0.50\pm0.59$ &$-7.06\pm2.1$ & &$-5.97\pm1.55$ &$-9.36\pm2.44$ \\
  \midrule{2-11}
  &10 &2 &$-2.76\pm0.83$ &$0.86\pm2.16$ & &$-3.43\pm0.63$ &$-6.17\pm1.68$ & &$-2.51\pm0.6$ &$-5.15\pm1.26$ \\
  & &5 &$11.64\pm4.55$ &$8.44\pm3.69$ & &$4.08\pm1.22$ &$-4.21\pm2.01$ & &$2.82\pm1.04$ &$-10.39\pm2.61$ \\
  & &10 &$101.82\pm25.59$ &$57.50\pm21.36$ & &$62.27\pm14.49$ &$6.65\pm11.63$ & &$42.13\pm10.48$ &$-29.69\pm12.25$ \\
  \bottomrule
\end{longtblr}
}

\subsection{Real Data Analysis}
\label{sec:results:real}
As discussed in the simulation studies, the generalized degrees of freedom (GDF) and the effective number of parameters estimated via cross-validation ($\hat{p}_{CV}$) may diverge when the fitted model is misspecified. In contrast, when the model provides an adequate description of the data-generating mechanism, these two measures tend to be closer, reflecting a more coherent assessment of model complexity relative to predictive performance. We investigate this phenomenon further through a real data analysis.

Our analysis proceeds as follows. First, we perform model selection using log-likelihood-based cross-validation $\ell_{CV}$ defined in Equation~(\ref{eq:cv}), which is asymptotically equivalent to the AIC (see Equation~(\ref{eq:AIC:lcv})), to identify the model with the smallest cross-validated negative log-likelihood ($-\ell_{CV}$). Second, we compare $\widehat{\mathrm{GDF}}$ and $\hat{p}_{CV}$ for the selected model, the intercept-only model, and the full model, in order to assess how these measures behave in practice under varying degrees of model adequacy.

The real data example considered here is the low birth weight dataset~\cite{hosmer2013applied}, which is available in the \texttt{aplore3 R} package. The dataset consists of 189 observations ($n=189$) on 11 variables, summarized in Table~\ref{tab:dataset}. The second variable, \texttt{low}, serves as the binary response, while the third through tenth variables (\texttt{age}, \texttt{lwt}, \texttt{race}, \texttt{smoke}, \texttt{ptl}, \texttt{ht}, \texttt{ui}, \texttt{ftv}) are treated as candidate predictors in the model fitting.
\begin{longtblr}[
    caption={Variables in low birth weight dataset.},
    label={tab:dataset}
  ]{
    vlines,
    colspec={*{3}{Q[l]}},
  }
  \toprule
  No. &Name &Description \\
  \midrule
  1 &\texttt{id} &Identification code \\
  2 &\texttt{low} &Low birth weight indicator (1: $\ge$2500 g, 2: $<$ 2500 g) \\
  3 &\texttt{age} &Age of mother (in Years) \\
  4 &\texttt{lwt} &Weight of mother at last menstrual period (in Pounds) \\
  5 &\texttt{race} &Race of mother (1: White, 2: Black, 3: Other) \\
  6 &\texttt{smoke} &Smoking statuse during pregnancy (1: No, 2: Yes) \\
  7 &\texttt{ptl} &History of premature labor (1: None, 2: One, 3: Two, etc) \\
  8 &\texttt{ht} &History of hypertension (1: No, 2: Yes) \\
  9 &\texttt{ui} &Presence of uterine irritability (1: No, 2: Yes) \\
  10 &\texttt{ftv} &Number of physician visits during the first trimester (1: None, 2: One, 3: Two, etc) \\
  11 &\texttt{bwt} &Recorded birth weight (in Grams) \\
  \bottomrule
\end{longtblr}


We consider all possible subsets of these covariates and perform model selection by minimizing the cross-validated negative log-likelihood ($-\ell_{CV}$). For each combination of the number of hidden units $H$ and decay parameter $\lambda$, the best model is identified as the one achieving the smallest value of $-\ell_{CV}$. The resulting best models are listed in Table~\ref{tab:best:model}.  
\begin{table}[htpb]
  \centering
  \caption{Best models for each pair of $\brc{H,\lambda}$ selected by least $-\ell_{CV}$.}
  \begin{tabular}{c*{3}{l}}
    \toprule
    $H$ &$\lambda=0.01$ &$\lambda=0.05$ &$\lambda=0.1$ \\
    \midrule
    2 &\texttt{lwt+ptl+ht} &\texttt{lwt+ptl+ht} &\texttt{lwt+ptl+ht} \\
    5 &\texttt{lwt+ptl} &\texttt{lwt+ptl+ht} &\texttt{lwt+ptl+ht} \\
    10 &\texttt{lwt+ptl} &\texttt{lwt+ptl+ht} &\texttt{lwt+ptl+ht} \\
    \bottomrule
  \end{tabular}
  \label{tab:best:model}
\end{table}

Under these selected models, we compute $\hat{p}_{CV}$ and $\widehat{\mathrm{GDF}}$, and compare them with the corresponding quantities obtained under the intercept-only model and the full model with same settings of $H$ and $\lambda$. The results of these comparisons are summarized in Table~\ref{tab:cmp:pcv:gdf}, where the results under the best models are highlighted. In the table, the symbols ``B'', ``I'', and ``F'' indicate the best, intercept-only, and full models, respectively.
{\scriptsize
\begin{longtblr}[
    caption={Comparison of cross-validated negative log-likelihood ($-\ell_{CV}$), effective number of parameters estimated via cross-validation ($\hat{p}_{CV}$), and generalized degrees of freedom ($\widehat{\mathrm{GDF}}$) for the best model (B) selected by minimizing $-\ell_{CV}$, the intercept-only model (I), and the full model (F), across different network configurations in the low birth weight dataset.},
    label={tab:cmp:pcv:gdf}
  ]{
    colspec={*{2}{Q[c]}*{3}{Q[r]}Q[c]*{3}{Q[r]}Q[c]*{3}{Q[r]}},
    cell{3}{3-13}={bg=gray9},
    cell{6}{3-13}={bg=gray9},
    cell{9}{3-13}={bg=gray9}
  }
  \toprule
  \SetCell[r=2]{c}$H$ &\SetCell[r=2]{c}model &\SetCell[c=3]{c}$\lambda=0.01$ & & & &\SetCell[c=3]{c}$\lambda=0.05$ & & & &\SetCell[c=3]{c}$\lambda=0.1$ \\
  \cmidrule{3-5}\cmidrule{7-9}\cmidrule{11-13}
  & &$-\ell_{CV}$ &$\hat{p}_{CV}$ &$\widehat{\mathrm{GDF}}$ & &$-\ell_{CV}$ &$\hat{p}_{CV}$ &$\widehat{\mathrm{GDF}}$ & &$-\ell_{CV}$ &$\hat{p}_{CV}$ &$\widehat{\mathrm{GDF}}$ \\
  \midrule
  2 &B &$110.2(3.4)$ &$11.7(3.4)$ &$8.7(0.1)$ & &$107.1(1.4)$ &$5.6(1.4)$ &$6.1(0.1)$ & &$107.5(1.0)$ &$5.1(1.0)$ &$4.9(0.1)$ \\
  &I &$117.3(0.0)$ &$-0.1(0.0)$ &$1.0(0.0)$ & &$117.3(0.0)$ &$0.0(0.0)$ &$0.9(0.0)$ & &$117.3(0.0)$ &$0.0(0.0)$ &$0.9(0.0)$ \\
  &F &$143.8(11.3)$ &$61.2(11.3)$ &$31.2(0.4)$ & &$121.4(5.1)$ &$33.9(5.1)$ &$25.6(0.3)$ & &$114.6(3.6)$ &$25.2(3.6)$ &$20.2(0.2)$ \\
  5 &B &$111.5(2.1)$ &$6.8(2.1)$ &$8.7(0.1)$ & &$107.6(1.4)$ &$6.1(1.4)$ &$6.8(0.1)$ & &$108.3(1.3)$ &$5.7(1.3)$ &$5.3(0.1)$ \\
  &I &$117.3(0.0)$ &$0.0(0.0)$ &$0.9(0.0)$ & &$117.3(0.0)$ &$0.0(0.0)$ &$0.9(0.0)$ & &$117.3(0.0)$ &$0.0(0.0)$ &$0.9(0.0)$ \\
  &F &$228.2(24.1)$ &$170.1(24.1)$ &$74.4(0.7)$ & &$153.3(10.8)$ &$84.1(10.8)$ &$56.1(0.5)$ & &$130.5(5.8)$ &$58.4(5.8)$ &$42.4(0.3)$ \\
  10 &B &$111.6(2.3)$ &$8.8(2.3)$ &$9.3(0.1)$ & &$107.8(1.4)$ &$6.6(1.4)$ &$6.8(0.1)$ & &$108.3(1.1)$ &$5.6(1.1)$ &$5.3(0.1)$ \\
  &I &$117.3(0.0)$ &$0.0(0.0)$ &$0.9(0.0)$ & &$117.3(0.0)$ &$0.0(0.0)$ &$0.9(0.0)$ & &$117.3(0.0)$ &$0.0(0.0)$ &$1.0(0.0)$ \\
  &F &$293.1(25.0)$ &$270.2(25.0)$ &$124.6(0.6)$ & &$173.1(10.4)$ &$70.5(10.4)$ &$86.7(0.3)$ & &$138.2(6.2)$ &$75.1(6.2)$ &$61.1(0.3)$ \\
  \bottomrule
\end{longtblr}
}


The results in Table~\ref{tab:cmp:pcv:gdf} show a clear contrast in the behavior of $\widehat{\mathrm{GDF}}$ and $\hat{p}_{CV}$ across different model choices. Under the best models selected by minimizing the cross-validated negative log-likelihood, the two quantities are close to each other, with their variability ranges overlapping when accounting for standard errors. This indicates a coherent assessment of model complexity and predictive performance when the model is well supported by the data.

In contrast, for the intercept-only and full models, $\widehat{\mathrm{GDF}}$ and $\hat{p}_{CV}$ differ substantially, with no overlap in their corresponding variability ranges. These discrepancies suggest that, when the model is either overly simplistic or overly complex relative to the data, the sensitivity-based measure of complexity (GDF) and the cross-validation-based measure ($\hat{p}_{CV}$) capture different aspects of model behavior.
We note that, in one intercept-only configuration, the estimated $\hat{p}_{CV}$ takes a slightly negative value. Since the intercept-only model has effectively zero complexity beyond the mean structure, $\hat{p}_{CV}$ is expected to be close to zero. The observed negative value is therefore attributed to numerical fluctuations arising from finite-sample variability and Monte Carlo error in cross-validation, rather than reflecting a meaningful negative model complexity.

\section{Conclusions}
\label{sec:conclusions}
In this work, we investigated the generalized degrees of freedom (GDF) as a measure of model complexity for feedforward neural networks, and compared it with likelihood-based alternatives, including the effective number of parameters estimated via cross-validation ($\hat{p}_{CV}$) and null degrees of freedom derived from likelihood ratio statistics. Through extensive simulation studies and a real data analysis, we highlighted fundamental differences in how these quantities behave under model misspecification, varying model structures and regularization strengths.

A key finding is that GDF provides a conceptually and practically robust measure of model complexity that can be computed regardless of whether the fitted model is correctly specified. Because GDF is defined through the sensitivity of fitted values to perturbations in the observed responses, it captures model flexibility and data adaptivity without relying on likelihood correctness or distributional assumptions. This property makes GDF particularly appealing for complex and flexible models, such as neural networks, where model misspecification is often unavoidable.

In contrast, likelihood-based measures such as $\hat{p}_{CV}$ and null degrees of freedom exhibit stronger dependence on the adequacy of the assumed model. When the model is well aligned with the data-generating machanism, these measures tend to agree more closely with GDF. However, under misspecification, substantial discrepancies can arise, and likelihood-based quantities may become unstable or lose their theoretical justification. Our results illustrate this limitation clearly in both simulation and real data settings.

Another practical advantage of GDF observed in our studies is its comparatively smaller variability relative to $\hat{p}_{CV}$. The cross-validation-based estimate $\hat{p}_{CV}$ relies on repeated data splitting and refitting, which introduces additional sources of variability. While this variability may be mitigated to some extent by increasing Monte Carlo replications or adjusting the number of folds, it remains inherent to the cross-validation procedure. In contrast, GDF estimates, particularly those obtained via the horizontal method, exhibit greater numerical stability across a wide range of configurations.

Taken together, these findings suggest that GDF serves as a reliable and broadly applicable measure of model complexity, especially in settings involving flexible models and potential model misspecification. Likelihood-based measures such as $\hat{p}_{CV}$ remain valuable when the modeling assumptions are approximately correct and predictive performance is the primary focus, but GDF offers a complementary perspective that emphasizes intrinsic model flexibility and robustness. Future work could explore principled strategies for stabilizing cross-validation-based complexity estimates and for combining sensitivity-based and likelihood-based measures into unified model assessment criteria.

\vspace{2em}
\declare{Author contributions}{The authors confirm contribution to the paper as follows: study conception and design: Zhou J, Landsittel D; analysis and interpretation of results: Zhou J, Landsittel D; draft manuscript preparation: Zhou J. All authors reviewed the results and approved the final version of the manuscript.}

\declare{Funding}{This research received no external funding.}

\declare{Conflict of interest}{The authors declare that they have no conflict of interest.}

\declare{Data availability}{All simulation studies and data analyses reported in this article were conducted using custom \texttt{R} code developed by the authors. The code used to generate the results presented in the figures and tables is permanently archived on \textbf{Zenodo} at \url{https://doi.org/10.5281/zenodo.18636066}. The development version and future updates are available at the \textbf{GitHub} repository: \\
\href{https://github.com/pmdnticc/neural_network_complexity}{https://github.com/pmdnticc/neural\_network\_complexity}.}

\bibliographystyle{unsrt}
\bibliography{references}

\end{document}